\documentclass[aps,prl,twocolumn,showpacs,superscriptaddress,groupedaddress,nofootinbib,preprintnumbers]{revtex4-1}
\pdfoutput=1
\usepackage{amsmath}
\usepackage{amsfonts}
\usepackage{amssymb}
\usepackage{graphicx, rotating}
\usepackage{epstopdf}
\usepackage{epsfig}
\usepackage{latexsym}
\usepackage{multirow}
\usepackage{color}
\usepackage{amsmath,amssymb}
\usepackage{slashed}
\usepackage[colorlinks=true,linkcolor=red,anchorcolor=black,citecolor=blue,filecolor=cyan,menucolor=red,runcolor=filecolor,urlcolor=blue,bookmarks=true,bookmarksnumbered=true]{hyperref}
\definecolor{rossos}{cmyk}{0,1,1,0.55}
\definecolor{bluscuro}{rgb}{0.15, 0.2, .85}
\definecolor{bluchiaro}{cmyk}{1,.3,0.,0.1}

\definecolor{rossos}{cmyk}{0,1,1,0.55}
\definecolor{bluscuro}{rgb}{0.15, 0.2, .85}
\definecolor{bluchiaro}{cmyk}{1,.3,0.,0.1}

\newcommand{\bc}{\begin{center}}
\newcommand{\ec}{\end{center}}
\newcommand{\pMET}{{\bf p}\llap{/\kern1.5pt}_T}
\newcommand{\bea}{\begin{eqnarray}}
\newcommand{\eea}{\end{eqnarray}}
\newcommand{\ignore}[1]{}
\def\BR{{\textrm{BR}}}
\def\stBR{\sigma\hspace{-1pt}\times\hspace{-1pt}\BR}
\newcommand{\be}{\begin{equation}}
\newcommand{\ee}{\end{equation}}

\def\({\left(}
\def\){\right)}
\def\<{\langle}
\def\>{\rangle}

\def\be{\begin{equation}}
\def\ee{\end{equation}}
\def\bry{\begin{array}}
\def\ery{\end{array}}
\def\bes{\begin{subequations}}
\def\ees{\end{subequations}}
\def\bit{\begin{itemize}}
\def\eit{\end{itemize}}
\def\ben{\begin{enumerate}}
\def\een{\end{enumerate}}

\newcommand{\MET}{E\llap{/\kern1.5pt}_T}
\definecolor{grey}{rgb}{0.6,0.6,0.6}
\definecolor{fuchsia}{rgb}{1,0,1}

\def\eWWWW{\epsilon_{\text{\tiny $W\hspace{-1pt}W\hspace{-2pt}\rightarrow\hspace{-2pt} W\hspace{-1pt}W$}}}
\def\eWZWW{\epsilon_{\text{\tiny $W\hspace{-1pt}Z\hspace{-2pt}\rightarrow\hspace{-2pt} W\hspace{-1pt}W$}}}
\def\eWWWZ{\epsilon_{\text{\tiny $W\hspace{-1pt}W\hspace{-2pt}\rightarrow\hspace{-2pt} W\hspace{-1pt}Z$}}}
\def\eWZWZ{\epsilon_{\text{\tiny $W\hspace{-1pt}Z\hspace{-2pt}\rightarrow\hspace{-2pt} W\hspace{-1pt}Z$}}}
\def\eWWZZ{\epsilon_{\text{\tiny $W\hspace{-1pt}W\hspace{-2pt}\rightarrow\hspace{-2pt} Z\hspace{-1pt}Z$}}}
\def\eWZZZ{\epsilon_{\text{\tiny $W\hspace{-1pt}Z\hspace{-2pt}\rightarrow\hspace{-2pt} Z\hspace{-1pt}Z$}}}
\def\SWZ{S_{\text{\tiny $W\hspace{-1pt}Z$}}}
\def\SWW{S_{\text{\tiny $W\hspace{-1pt}W$}}}
\def\SZZ{S_{\text{\tiny $Z\hspace{-1pt}Z$}}}
\begin{document}

\preprint{DFPD-2015/TH/16}
\preprint{MITP/15-044}

\title{A composite Heavy Vector Triplet in the ATLAS di-boson excess}

%\date{\today}
%

\author{Andrea Thamm}
%\email{ferretti@chalmers.se}
\address{PRISMA Cluster of Excellence \& Mainz Institute for Theoretical Physics,\\
Johannes Gutenberg University, 55099 Mainz, Germany}
\author{Riccardo Torre and Andrea Wulzer}
%\email{riccardo.torre@pd.infn.it}
\address{Dipartimento di Fisica e Astronomia, Universit\`a di Padova, and INFN Sezione di Padova, Italy}
\begin{abstract}
Composite vector resonances in the triplet of the SM SU$(2)_{L}$ gauge group are a universal prediction of ``natural'' new physics models involving a new strongly-interacting sector and are therefore among the most plausible new particles that the LHC could discover. We consider the possibility that one such triplet could account for the ATLAS excess in the invariant-mass spectrum of boson-tagged jets and we assess the compatibility of this hypothesis with all other relevant resonance searches. 
We find that the hypothesis is not excluded and that the predicted signal is close to the expected sensitivity of several channels, some of which show an upper fluctuation of the observed limit while others do not. An accurate study of the signal compatibility with these fluctuations could only be performed by the ATLAS and CMS collaborations. 
\end{abstract}
\pacs{12.60.Fr, 12.60.Rc} 
\keywords{}

\maketitle
%%%%%%%%%%%%%%%%%%%%%%%%%%%%%%%%%%%%%%%
%\setcounter{page}{2}
%%%%%%%%%%%%%%%%%%%%%%%%%%%%%%%%%%%%%%%

\noindent{\bf Introduction:} 

New physics scenarios addressing the Higgs mass Hierarchy Problem fall into only two categories. There are those that rely on supersymmetry and those that advocate the existence of a new strongly-interacting composite sector which is ultimately responsible for Electro-Weak Symmetry Breaking (EWSB). Models where the Higgs field emerges as a composite pseudo-Nambu--Goldstone boson are particularly plausible realisations of the composite-sector EWSB paradigm. Spin one vector resonances with EW quantum numbers and TeV-sized mass are unavoidably present in this class of models, since they are associated with the \mbox{SU(2)$_L\times$U$(1)_Y$} EW gauge symmetries and to the corresponding global current operators. EW-charged vector particles are thus some of the most theoretically well motivated resonances that the LHC could discover. In what follows we will focus on vector resonances in the triplet of the EW  \mbox{SU(2)$_L$} group and with vanishing hypercharge, namely the ones associated with the \mbox{SU(2)$_L$} current. 

Composite Heavy Vector Triplets (HVT) are rather elusive particles. In ref.~\cite{Pappadopulo:2014qza} we emphasised that a large fraction of the ``natural'' parameter space of composite HVT was still poorly explored by the $8$\,TeV LHC searches available at that time. Specifically, a triplet with a mass of $2$\,TeV which complies well with expectations based on naturalness and is also marginally compatible with the indirect constraints from EW Precision Tests (EWPT), is still viable in the composite case (represented by ``Model B'' of ref.~\cite{Pappadopulo:2014qza}) while it is excluded if the triplet has an ``elementary'' origin (corresponding to ``Model A'') from a weakly-coupled gauge theory. We also identified a critical value, $g_V\simeq3$, of the intrinsic HVT coupling for which the $2$\,TeV mass hypothesis was close to being tested by collider searches in several different final states, including leptons, hadronically or semileptonically decaying EW boson pairs and $WH$ or $ZH$ channels involving a Higgs boson.

This suggests that the $2$\,TeV excess recently reported by the ATLAS collaboration in the search for hadronically decaying EW bosons \cite{Aad:2015owa} might fit well with a composite HVT close to the critical mass-coupling configuration. The sensitivity of the analysis might have improved just enough to make the critical configuration show up, though with a still limited statistical significance. The possibility of explaining the ATLAS excess by composite vector resonances has been recently discussed in refs.~\cite{Fukano:2015hga,Franzosi:2015zra} and in ref.~\cite{Aguilar-Saavedra:2015rna}. In particular this last article claims that the HVT interpretation is excluded by other search channels, but it lacks a quantitative assessment of the signal cross-section needed to reproduce the excess, which is essential for quantitative comparisons of the putative signal with the other searches. Performing this quantitative assessment and comparison is the aim of the present article.\\ \ \\ 
\noindent{\bf The model:} 

The reader is referred to ref.~\cite{Pappadopulo:2014qza} for a detailed description of the HVT setup and phenomenology, here we only recall the few basic qualitative features that are needed to understand the results that follow.\footnote{Many of the statements below are subject  to small corrections, which are irrelevant for the present discussion but could become crucial for an accurate comparison of the model with data.} First, an HVT contains one electrically charged $V^\pm$ and one neutral $V^0$ state. The two states are essentially degenerate since a mass splitting only comes from EWSB effects, which are tiny for a $2$\,TeV particle. The relative magnitude of the $V^\pm$ and $V^0$ couplings to quarks are fixed by the \mbox{SU(2)$_L$} symmetry and thus the $V^\pm$ and $V^0$ relative production rates are entirely determined by the parton luminosities. For all masses considered here the $V^\pm$ production rate is approximately twice the one of the $V^0$. The overall strength of the quark (and lepton) couplings are fixed, in the notation of ref.~\cite{Pappadopulo:2014qza}, in terms of the $g_V$ coupling, and they scale as $g^2/g_V$, where $g$ is the \mbox{SU(2)$_L$} SM coupling. This leads to a rather peculiar situation in which a large composite sector coupling $g_V$ corresponds to a small production rate. Also, a large $g_V$ makes the branching ratio to fermionic final states small. In the two lepton channels, this is of the order of a few permille for the value $g_V\simeq3$ we consider here. The HVT thus decays primarily to the EW bosons and to the Higgs particle, with relative branching ratios fixed by symmetries to be all equal in the allowed decay channels, namely
\begin{eqnarray}
&&\BR(V^\pm\rightarrow W^\pm Z )=\BR(V^\pm\rightarrow W^\pm h )\nonumber\\
&&=\BR(V^0\rightarrow W^+ W^- )=\BR(V^0\rightarrow Z h )\simeq\frac12\,.
\end{eqnarray}

These considerations clearly illustrate how tightly constrained the HVT phenomenology is. In particular all the di-boson final state processes are described, in full generality, by one single coupling $g_V$ which controls the production rate. Fixing the latter to reproduce the ATLAS excess will allow us to make sharp predictions for the other search channels and to check the compatibility of the signal hypothesis. In what follows, if not otherwise specified, we will consider the reference Model B of ref.~\cite{Pappadopulo:2014qza}, where the branching ratios to opposite-sign di-leptons and lepton-neutrino final states are also fixed by $g_V$. However a second order one parameter $a_\rho$, which is set to one in Model~B, is also present for a generic composite HVT  \cite{Pappadopulo:2014qza,Contino:2011np}. This parameter allows us to change the relative branching ratios of bosonic and fermionic final states and will become relevant when we compare our signal with experimental searches in the di-lepton and lepton-neutrino final states. All the numerical results that follow are obtained by the web tool in \cite{projectwebpage}. \\ \ \\ 
{\bf Fitting the excess:} 

In ref.~\cite{Aad:2015owa} the ATLAS collaboration reported a statistically significant (up to $3.4\sigma$ local translating into $2.5\sigma$ global significance in the $WZ$ channel) excess in the invariant mass distribution of two jets, tagged as hadronically decaying $W$ or $Z$ EW bosons by employing jet substructure techniques. The analysis has some discriminating power on the nature of the hadronically decaying EW bosons, which are distinguished by the different invariant mass of the single boson-tagged jet. Therefore three different event selections are considered, corresponding to different selected windows for the two fat-jet invariant masses. Three di-jet invariant mass distributions are reported for the $WZ$, $WW$ and $ZZ$ selections. However the selection regions largely overlap, implying that the three channels are correlated and by far not independent.

Given that the $V^\pm$ component of the HVT, which has the largest production rate, decays to $W Z$, and that this is the channel with the largest excess, we start our analysis from the $WZ$ selection results, reported in figure~5~(a) of ref.~\cite{Aad:2015owa}. In particular we focus on the $5$ bins that span the invariant mass range from $1.75$ to $2.25$\,TeV di-jet invariant mass, where a total of $20$ events have been counted with a SM-expected of $13$ and thus an excess of $7$ events. Obviously this way of looking at the excess by integrating over a rather large window does not do justice to its statistical significance, which is dominated by the much more pronounced discrepancy with the background in a single bin centred at $2$\,TeV. However it is the most convenient approach for the determination of the signal cross-section that could have produced it. 

We thus need $7$ signal events in the $[1.75,2.25]$\,TeV mass window. We can get a first estimate of the required cross-section by looking again at figure~5~(a) of ref.~\cite{Aad:2015owa}, which also reports the expected distribution of a $2$\,TeV $W'\rightarrow WZ$ signal with a cross-section times branching-ratio ($\stBR$) of $3.17$\,fb. This signal would lead to $3.4$ events in the mass window we consider, thus to produce the excess we need $7/3.4$ more events and thus a cross-section of $6.5$\,fb. By looking at the final exclusion plot of ref.~\cite{Aad:2015owa} for the $WZ$ channel one might find this result surprising, given that this cross section is not only considerably below the excluded one (about $40$\,fb at $2$\,TeV), but also below the expected limit of around $10$\,fb. However the way in which a weak signal leading to few events in excess affects the exclusion limit and makes it jump above the expected is a statistical question that depends on the number of SM background events and on the systematic uncertainty of the signal acceptance in a non-trivial way. Therefore there is a priori no contradiction and furthermore our estimate of the signal cross section seems unquestionable since it is based on a rescaling of the ATLAS simulation result.

Up to now we estimated the cross-section that we would need if our signal was entirely coming from the charged $V^\pm$ production, but the neutral HVT component $V^0$, decaying to $WW$, also contributes to the $WZ$-selected events because the selection cuts poorly distinguish between $W$ and $Z$ vector bosons as previously explained. We estimate the contamination of physical $WW$ events in the $WZ$-selected sample by proceeding as follows. ATLAS provides us with the invariant fat-jet mass distribution originating from a $W$ and a $Z$ bosons and reports the windows in this variable, constructed for the two fat-jets, which are used to select the different di-boson samples. %These windows are centred at 82.4 GeV and 92.8 GeV respectively for the $W$ and $Z$ boson and are $\pm 13 $ GeV wide. 
After properly taking into account the presence of a large overlap region between the $W$ and $Z$ boson selection windows (which is relevant for the $WZ$ selection), this allows us to compute the efficiencies of the jet invariant mass cuts that define the three signal regions for the $WZ$, $WW$ and $ZZ$ physical vector boson configurations. The result, including the branching ratios for hadronic decays, is
\begin{equation}
\left[\begin{array}{cc}
\eWWWW  & \eWZWW \\
\eWWWZ & \eWZWZ \\
\eWWZZ & \eWZZZ 
\end{array}
\right]=\left[\begin{array}{cc}
0.18 & 0.15 \\
0.17 & 0.21 \\
0.07 & 0.12
\end{array}
\right]\,.
\end{equation}
The prediction for the signal in the di-jet invariant mass window that we consider is thus
\be\label{eq2}
\SWZ\hspace{-2pt}=\hspace{-2pt}{\mathcal{L}}\,{\mathcal{A}}\left[(\stBR)_{V^\pm}\eWZWZ+(\stBR)_{V^0}\eWWWZ \right]\,,
\ee
where an ``acceptance'' factor ${\mathcal{A}}$ has been included to reproduce the impact of the other selection cuts, performed ``before'' the final selection on the fat-jets invariant masses whose efficiency we estimated above. The acceptance also includes the cut in the $[1.75,2.25]$\,TeV mass window that we used to define our signal region. Its value, which is around $0.25$ for $m_V=2$\,TeV, has been estimated by the $W'$ total selection efficiency (reported in fig.~2 (b) of ref.~\cite{Aad:2015owa}) and by the $W'$ signal shape in figure~5~(a) of ref.~\cite{Aad:2015owa}. The luminosity  ${\mathcal{L}}=20.3\,{\textrm{fb}}^{-1}$ is obviously also present in the equation above.

In our model, the charged and neutral $\stBR$ are controlled by the unique parameter $g_V$, besides the resonance mass $m_V$. By eq.~\eqref{eq2} and imposing $\SWZ=7$ we obtain, for $m_V=2$\,TeV, a determination of $g_V = 2.81$. Not surprisingly, this value is close to the ``critical'' point we discussed in the Introduction. The corresponding charged and neutral $\stBR$ are reported in Table \ref{Table:Predictions}. 
%\begin{table}[t]
%\begin{center}
%%%
%{
%\begin{tabular}{c|c|c}
%\hline
%\hline
%Mass [TeV] & Predicted $g_{V}$ & Signal $\stBR$ \\ 
%1.8 & & \multirow{3}{*}{$\bry{l}\dst (\stBR)_{V^\pm}= \text{ fb}\\\dst (\stBR)_{V^0}= \text{ fb}\ery$} \\ 
%1.9 & &  \\
%2.0 & $2.81^{+1.54}_{- 0.82} $ &  \\
%\hline
%\hline
%\end{tabular}
%}
%%%
%\\[0.1cm]
%\caption{\small Predicted values for the signal $\stBR$ and corresponding values of $g_{V}$ with $68\%$ confidence intervals (see the text for details).}\label{Table:Predictions}
%\end{center}
%\end{table}
\begin{table}[t]
\begin{center}
{
\begin{tabular}{c|c|c|c}
\hline
\hline
$m_V$ [TeV] & $g_V$ & $(\stBR)_{V^\pm}$ [fb] & $(\stBR)_{V^0}$ [fb]\\ 
\hline
$1.8$ & $3.95^{+1.65}_{-0.88}$ & $4.51$ & $2.04$ \\
$1.9$ & $3.37^{+1.63}_{-0.83}$ & $4.63$ & $2.09$ \\
$2.0$ & $2.81^{+1.54}_{-0.82}$ & $4.79$ & $2.16$ \\
\hline
\hline
\end{tabular}
}
\\[0.1cm]
\caption{\small The signal $\stBR$ and the corresponding values of $g_{V}$ with $68\%$ confidence intervals.}\label{Table:Predictions}
\end{center}
\end{table}
The determination of $g_V$ is obviously affected by a large error. By assuming Poisson statistics for the data counting, neglecting the error on the SM background fit and the one on the signal acceptance (which is actually considerable), central $68\%$~CL \cite{Beringer:1900zz} intervals on the signal\footnote{In the following we will refer to the central $68\%$~CL intervals as $1\sigma$ intervals.} are obtained and translated in $g_V$ error bars. The result is $g_V=2.81^{+1.54}_{- 0.82} $. 

This is for $m_V=2$\,TeV, but it is clear that the excess does not uniquely select this value of the mass. The peak of the distribution is indeed in the $2$\,TeV bin for the $WZ$ and $WW$ selection, but sits at $1.9$\,TeV for $ZZ$. Furthermore the histogram bins are $100$\,GeV wide and moreover ref.~\cite{Aad:2015owa} estimates a $2\%$ systematic uncertainty on the jet $p_T$ scale, which could easily shift the $2$\,TeV points by $40$\,GeV. Therefore we also consider other mass hypotheses, namely $m_V=1.8$\,TeV (which is probably too low) and $m_V=1.9$\,TeV. We determine the value of $g_V$ at these masses, and its error bars, as the one that gives the same total number of signal events, in the whole di-jet invariant mass range, that we would obtain with the $2$\,TeV resonance cross-section. A mild variation of the acceptance ${\mathcal{A}}$, that decreases with the mass, makes the required signal cross-section a bit smaller at the lower mass points.  The results, reported in Table \ref{Table:Predictions}, define a region in the $(m_V,g_V)$ plane where the HVT is supposed to sit.

Having determined the $(m_V,g_V)$ region by the $WZ$-selected events we should now check its compatibility with the $WW$- and $ZZ$-selected samples. The number of signal events can be predicted by the obvious generalisations of eq.~\eqref{eq2}, that gives (for $m_V=2$~TeV) $1\,\sigma$ intervals $\SWW\in [2.2, 10.3]$ and $\SZZ\in[1.4, 6.6]$ in the $[1.75,2.25]$\,TeV di-jet mass window. The observed excesses are respectively $4.2$ and $6.4$ events and are well compatible with the expectations. Given the limited statistics and the consequently large error bars and the large contamination between different channels, this result is an almost trivial check of our signal hypothesis.\\ \ \\ 
{\bf Other searches:} 

\begin{figure*}
\begin{center}
\includegraphics[scale=1.1]{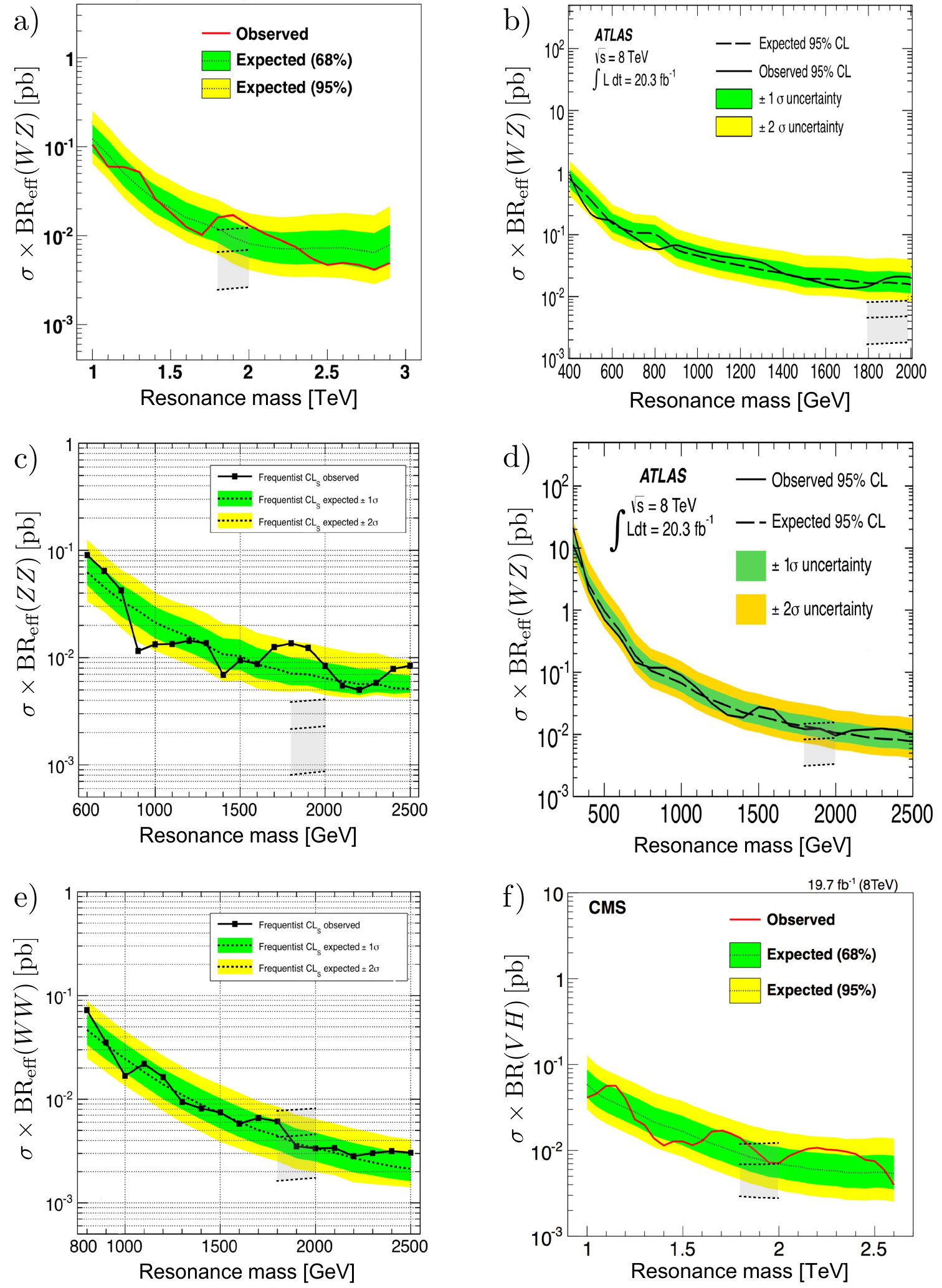}\hspace{0.5cm}
\caption{\small{The HVT fit of the ATLAS excess compared with other experimental searches.}}\label{results}
\end{center}
\end{figure*}

We now check the compatibility of our signal with other experimental searches, starting from the CMS counter-part of ref.~\cite{Aad:2015owa}, namely a resonance search in the invariant mass of boson-tagged di-jets \cite{CMSCollaboration:2014df}. This analysis reports separate $\stBR$ exclusion limits for different signal hypotheses, among which all the ones that produce mostly longitudinally polarised EW bosons (namely $W'$ and bulk graviton), like our model, could be useful to constrain our signal. We will only consider the $W'\rightarrow WZ$ interpretation, which is more similar to our signal, however we have checked that the others give identical results. In order to properly superimpose our signal with the CMS exclusion we need to take into account that we not only have a charged $V^\pm$ state, but also a neutral $V^0$ one that contributes in the search region, similarly to what we discussed above for the ATLAS search. In the CMS analysis no attempt is made to distinguish $W$ from $Z$ boson jets and the acceptance in the search region seems identical, as far as we can infer from ref.~\cite{CMSCollaboration:2014df}, for a physically hadronically decaying $WZ$ or $WW$ pair. Therefore we compare the CMS limit with an ``effective'' $\stBR$ obtained as the sum of the charged vector $\stBR$ and the one of the neutral vector, properly rescaled to take into account the (tiny) difference in the hadronic $W$ and $Z$ $\BR$s, namely
\be
(\stBR)_{\textrm{eff}}=(\stBR)_{V^\pm}+\frac{\BR_{W\rightarrow{\textrm{had}}}}{\BR_{Z\rightarrow{\textrm{had}}}}(\stBR)_{V^0}\,.
\ee
The result is reported in fig.~\ref{results}a). It not only shows that our signal is not excluded, but also that it could account for the small upper fluctuation of the limit. 

Let us now consider semi-leptonic di-boson searches, in which one of the EW bosons decays hadronically, and is seen as a single fat jet, while the other decays leptonically. The relevant searches are presented by ATLAS in refs.~\cite{Aad:2014xka,Aad:2015ufa} and by CMS in ref.~\cite{CMSCollaboration:2014ke}. All those analyses can not significantly discriminate between hadronically decaying $W$ and $Z$ bosons. Therefore in what follows we will assume, as for the fully hadronic CMS search, an equal probability for hadronic $W$ and $Z$ ending up in the search region. One leptonically decaying $Z$ boson was considered by ATLAS in ref.~\cite{Aad:2014xka} and $\stBR$ limits where set on a $W'\rightarrow WZ$ signal and on a bulk graviton decaying to $ZZ$. We consider the former one, which is more suited for our signal, but a very similar result would be obtained with the latter. Given that $V^0$ does not decay to $ZZ$, the only way to get a leptonic $Z$ from our HVT is by the production of the $V^\pm$ decaying to $WZ$. The $\stBR$ limit on the $W'\rightarrow WZ$ signal hypothesis can thus be directly used in our model, with the result reported in fig.~\ref{results}b). The corresponding CMS result, in ref.~\cite{CMSCollaboration:2014ke}, is a bit more difficult to use since it sets a limit on a bulk graviton decaying to $ZZ$. By assuming the same selection efficiency for the bulk graviton and for our signal (which is not necessarily accurate), the effective $\stBR$ to be employed for the comparison is
\be
(\stBR)_{\textrm{eff}}=\frac{\BR_{W\rightarrow{\textrm{had}}}}{2\,\BR_{Z\rightarrow{\textrm{had}}}}(\stBR)_{V^\pm}\,,
\ee
which takes into account the important combinatorial factor of $2$ that is present for the semileptonic decay of the $ZZ$ pair, produced by the bulk graviton, but is not present for our $WZ$ signal. The result is shown in fig.~\ref{results}c). We conclude that our signal is not excluded by leptonic $Z$ searches and that it could be actually responsible (at least in the upper side of the $1\sigma$ band, but this should be checked) for the CMS $2\sigma$ excess of ref.~\cite{CMSCollaboration:2014ke}.

We now turn to the case of one leptonic $W$, which was studied by ATLAS in ref.~\cite{Aad:2015ufa} and by CMS in ref.~\cite{CMSCollaboration:2014ke}. The ATLAS collaboration provides an interpretation of the search in terms of a $W'\rightarrow WZ$, which we can compare with our model by means of the effective $\stBR$
\be
(\stBR)_{\textrm{eff}}=(\stBR)_{V^\pm}+\frac{2\,\BR_{W\rightarrow{\textrm{had}}}}{\BR_{Z\rightarrow{\textrm{had}}}}(\stBR)_{V^0}\,.
\ee
Notice that the combinatorial factor of $2$ now sits in the numerator and enhances the neutral $V^0$ contribution. The result is displayed in figure fig.~\ref{results}d), which shows that the central value of our signal is not excluded while the upper part of the cross-section band is in tension with this analysis. The leptonic $W$ CMS analysis \cite{CMSCollaboration:2014ke} provides a limit interpretation in terms of a bulk graviton signal decaying to $WW$. We use it by assuming the same efficiency for the bulk graviton and our HVT, which is, as we want to stress once again, not necessarily a good approximation. For the comparison we use an effective $\stBR$, constructed along the same lines we described above, which takes into account the combined contribution of the neutral and the charged resonances. The limit is a bit stronger than the one we obtained from the ATLAS analysis and is shown in fig.~\ref{results}e).

We finally consider final states involving the Higgs particle. The $HZ$ channel with a hadronically decaying EW boson and $H \to \tau \tau$ was considered by CMS in ref.~\cite{Khachatryan:2015ywa}, while the $HW$ channel with fully hadronic decays was studied in ref.~\cite{Khachatryan:2015bma}. The combination with the $HZ$ channel and an interpretation in terms of the HVT model, which we can thus use directly, was also performed in ref.~\cite{Khachatryan:2015bma}, with the result shown in fig.~\ref{results}f). The central value of the signal is still below the expected limit, though once again a certain tension is present with the upper side of the $1\sigma$ band. A similar analysis has also been performed by ATLAS \cite{Aad:2015yza} with a leptonically decaying EW boson and $H\to b\bar{b}$ and by CMS with a leptonic W and $H\to b\bar{b}$ \cite{CMS:2015gla}. Our signal is not excluded by these searches, and moreover the latter displays a $2\sigma$ fluctuation of the limit which could perhaps be accounted for by our signal, but this requires further study. Finally, further analyses in the di-boson channel decaying fully leptonically have been performed by ATLAS \cite{ATLAScollaboration:2014uc} and CMS \cite{Khachatryan:2014xja}. We do not include them here because they are less sensitive.

Before concluding, it is worth discussing two-lepton final state searches, which include $l^{+}l^{-}$ from ATLAS \cite{ATLAScollaboration:2014eb} and CMS \cite{Khachatryan:2014fba} and the ATLAS \cite{ATLAScollaboration:2014gl} and CMS \cite{CMSCollaboration:2014vq} $l \nu$ searches. They have a considerably better reach on $\stBR$ than the di-boson ones and thus are the natural killers of generic interpretations of the ATLAS excess in terms of weakly-coupled vector resonances of elementary nature. In the Introduction we discussed that these constraints are naturally avoided by a composite HVT and indeed we checked that for the reference HVT model (Model~B of ref.~\cite{Pappadopulo:2014qza}) we considered here, the two-lepton exclusion limits are above our predictions. Namely, they are around one order of magnitude above for the central value of the cross-section and roughly comparable for the upper $1\sigma$ value. The two-lepton searches would become more relevant if the order one parameter $a_\rho$, which we discussed in the Introduction, was not exactly set to one as in Model B. By lowering $a_\rho$ we can boost the $\stBR$ in the di-lepton channel while leaving all the results on di-boson final states unchanged. For instance, considering $a_{\rho}=1/\sqrt{2}$ and taking $g_{V}$ slightly smaller than our best fit, i.e.~$g_{V}=2.2$, the $\stBR$ in the di-lepton final state is around one order of magnitude bigger, while leaving all the di-boson channels unchanged (see ref.~\cite{Pappadopulo:2014qza} for further details on the dependence of the various channels on the $a_{\rho}$ parameter). Such a choice of parameters could also account for the $2\sigma$ fluctuation in the CMS di-lepton analysis \cite{Khachatryan:2014fba}.\\ \ \\ 
{\bf Conclusions:} 

We studied the compatibility of the ATLAS excess of ref.~\cite{Aad:2015owa}, interpreted in terms of a composite HVT, with the other relevant experimental searches. We showed that this interpretation is not excluded and that it could also explain some less significant excesses in other channels, especially  if it sits on the upper side of the $1\sigma$ cross-section band. A certain tension with this latter possibility comes however from other searches. Broadly speaking, the fact that a weak signal shows up as an excess in some channels and not in others seems to us perfectly compatible with the large expected statistical fluctuations and with the uncorrelated systematic uncertainties which might boost the signal in some channel and deplete it in others. A combined analysis of the different searches would be needed for a fully quantitative assessment of the statistical viability of our signal hypothesis.

The ATLAS and CMS collaborations should study this signal for at least two reasons. First, a theoretically credible and motivated interpretation of the ATLAS excess, which is still far from being a discovery, would greatly increase the degree of belief that it could really be due to new physics. Exotic interpretations of the excess, like those that appeared recently in refs.~\cite{Fukano:2015hga,Hisano:2015gna,Cheung:2015nha,Dobrescu:2015qna,Alves:2015mua,Gao:2015irw}, are interesting but less helpful in that respect. Second, a combined study of different search channels, of course under the assumption that the right model has been selected for the combination, would enormously accelerate a discovery at the forthcoming LHC run. The composite HVT is such a well motivated and predictive framework that it is definitely worth trying.

Other motivated composite vector particles might also be considered, the simplest option being the custodial group $({\mathbf{1}},{\mathbf{3}})$ representation associated to the \mbox{SU$(2)_R$} global current. Actually, only the neutral component of this multiplet couples strongly to quarks and can be copiously produced at the LHC, leading effectively to a composite version of the neutral $Z'$. We expect its phenomenology to be quite similar to the one of the HVT, but we leave this to future investigations. Even more complicated multi-particle composite vector models could be considered, such as the one discussed in ref.~\cite{Greco:2014aza}.
\\ 
{\bf Acknowledgments:}

%\end{center}
%\vspace{-0.2cm}
We would like to thank A.\,Picazio and M.\,Pierini for discussions. A.T.~acknowledges support from the Cluster of Excellence {\it Precision Physics, Fundamental Interactions and Structure of Matter} (PRISMA -- EXC 1098). The work of R.T. was supported by the Italian PRIN {no.}~2010YJ2NYW$\_$003. A.W.~acknowledges the MIUR-FIRB grant \sloppy\mbox{RBFR12H1MW} and the ERC Advanced Grant no.267985 ({\emph{DaMeSyFla}}).

%\appendix

%\section{Appendix}

\bibliographystyle{mine}
\bibliography{bibliography}

\providecommand{\href}[2]{#2}\begingroup\raggedright\begin{thebibliography}{10}

\bibitem{Pappadopulo:2014qza}
D.~Pappadopulo, A.~Thamm, R.~Torre, and A.~Wulzer,
  \href{http://dx.doi.org/10.1007/JHEP09(2014)060}{{\em JHEP} {\bfseries 1409}
  (2014) 060}, \href{http://arxiv.org/abs/1402.4431}{{\tt arXiv:1402.4431}} [\href{http://inspirehep.net/record/1281686}{Inspire}].
%%CITATION = ARXIV:1402.4431;%%.

\bibitem{Aad:2015owa}
{\bfseries ATLAS} Collaboration, G.~Aad {\em et al.,}
  \href{http://arxiv.org/abs/1506.00962}{{\tt arXiv:1506.00962}} [\href{http://inspirehep.net/record/1374218}{Inspire}].

\bibitem{Fukano:2015hga}
H.~S. Fukano, M.~Kurachi, S.~Matsuzaki, K.~Terashi, and K.~Yamawaki,
  \href{http://arxiv.org/abs/1506.03751}{{\tt arXiv:1506.03751}} [\href{http://inspirehep.net/record/1375823}{Inspire}].

\bibitem{Franzosi:2015zra}
D.~B. Franzosi, M.~T. Frandsen, and F.~Sannino,
  \href{http://arxiv.org/abs/1506.04392}{{\tt arXiv:1506.04392}} [\href{http://inspirehep.net/record/1376127}{Inspire}].

\bibitem{Aguilar-Saavedra:2015rna}
J.~Aguilar-Saavedra, \href{http://arxiv.org/abs/1506.06739}{{\tt
  arXiv:1506.06739}} [\href{http://inspirehep.net/record/1377367}{Inspire}].

\bibitem{Contino:2011np}
R.~Contino, D.~Marzocca, D.~Pappadopulo, and R.~Rattazzi,
  \href{http://dx.doi.org/10.1007/JHEP10(2011)081}{{\em JHEP} {\bfseries 10}
  (2011) 081}, \href{http://arxiv.org/abs/1109.1570}{{\tt arXiv:1109.1570}} [\href{http://inspirehep.net/record/926810}{Inspire}].

\bibitem{projectwebpage}
D.~Pappadopulo, A.~Thamm, R.~Torre, and A.~Wulzer,
  \href{http://heidi.pd.infn.it/html/vector/index.html}{Webpage}.

\bibitem{Beringer:1900zz}
{\bfseries PDG} Collaboration, J.~Beringer {\em et al.,}
  \href{http://dx.doi.org/10.1103/PhysRevD.86.010001}{{\em Phys. Rev.}
  {\bfseries D 86} (2012) 010001} [\href{http://inspirehep.net/record/1126428}{Inspire}].

\bibitem{CMSCollaboration:2014df}
{\bfseries CMS} Collaboration, V.~Khachatryan
  \href{http://dx.doi.org/10.1007/JHEP08(2014)173}{{\em JHEP} {\bfseries 08}
  (2014) 173}, \href{http://arxiv.org/abs/1405.1994}{{\tt arXiv:1405.1994}} [\href{http://inspirehep.net/record/1294937}{Inspire}].

\bibitem{Aad:2014xka}
{\bfseries ATLAS} Collaboration, G.~Aad {\em et al.,}
  \href{http://dx.doi.org/10.1140/epjc/s10052-015-3261-8}{{\em Eur.Phys.J.}
  {\bfseries C75} (2015) 69}, \href{http://arxiv.org/abs/1409.6190}{{\tt
  arXiv:1409.6190}} [\href{http://inspirehep.net/record/1318483}{Inspire}].

\bibitem{Aad:2015ufa}
{\bfseries ATLAS} Collaboration, G.~Aad {\em et al.,}
  \href{http://dx.doi.org/10.1140/epjc/s10052-015-3425-6}{{\em Eur.Phys.J.}
  {\bfseries 5} (2015) 209}, \href{http://arxiv.org/abs/1503.04677}{{\tt
  arXiv:1503.04677}} [\href{http://inspirehep.net/record/1352826}{Inspire}].

\bibitem{CMSCollaboration:2014ke}
{\bfseries CMS} Collaboration, V.~Khachatryan
  \href{http://dx.doi.org/10.1007/JHEP08(2014)174}{{\em JHEP} {\bfseries 08}
  (2014) 174}, \href{http://arxiv.org/abs/1405.3447}{{\tt arXiv:1405.3447}} [\href{http://inspirehep.net/record/1296080}{Inspire}].

\bibitem{Khachatryan:2015ywa}
{\bfseries CMS} Collaboration, V.~Khachatryan {\em et al.,}
  \href{http://arxiv.org/abs/1502.04994}{{\tt arXiv:1502.04994}} [\href{http://inspirehep.net/record/1345160}{Inspire}].

\bibitem{Khachatryan:2015bma}
{\bfseries CMS} Collaboration, V.~Khachatryan {\em et al.,}
  \href{http://arxiv.org/abs/1506.01443}{{\tt arXiv:1506.01443}} [\href{http://inspirehep.net/record/1374621}{Inspire}].

\bibitem{Aad:2015yza}
{\bfseries ATLAS} Collaboration, G.~Aad {\em et al.,}
  \href{http://dx.doi.org/10.1140/epjc/s10052-015-3474-x}{{\em Eur.Phys.J.}
  {\bfseries C75} (2015) 263}, \href{http://arxiv.org/abs/1503.08089}{{\tt
  arXiv:1503.08089}} [\href{http://inspirehep.net/record/1356730}{Inspire}].

\bibitem{CMS:2015gla}
{\bfseries CMS} Collaboration {\em CMS Note} {\bfseries
  \href{http://cds.cern.ch/record/2002903/files/EXO-14-010-pas.pdf}{CMS-PAS-EXO-14-010}}
  (2015) [\href{http://inspirehep.net/record/1356348}{Inspire}].

\bibitem{ATLAScollaboration:2014uc}
{\bfseries ATLAS} Collaboration, G.~Aad {\em et al.,}
  \href{http://dx.doi.org/10.1016/j.physletb.2014.08.039}{{\em Phys. Lett.}
  {\bfseries B 737} (2014) 223--243},
  \href{http://arxiv.org/abs/1406.4456}{{\tt arXiv:1406.4456}} [\href{http://inspirehep.net/record/1300821}{Inspire}].

\bibitem{Khachatryan:2014xja}
{\bfseries CMS Collaboration} Collaboration, V.~Khachatryan {\em et al.,}
  \href{http://arxiv.org/abs/1407.3476}{{\tt arXiv:1407.3476}} [\href{http://inspirehep.net/record/1306289}{Inspire}].

\bibitem{ATLAScollaboration:2014eb}
{\bfseries ATLAS} Collaboration, G.~Aad {\em et al.,}
  \href{http://dx.doi.org/10.1103/PhysRevD.90.052005}{{\em Phys. Rev.}
  {\bfseries D 90} (2014) 052005}, \href{http://arxiv.org/abs/1405.4123}{{\tt
  arXiv:1405.4123}} [\href{http://inspirehep.net/record/1296830}{Inspire}].

\bibitem{Khachatryan:2014fba}
{\bfseries CMS} Collaboration, V.~Khachatryan {\em et al.,}
  \href{http://dx.doi.org/10.1007/JHEP04(2015)025}{{\em JHEP} {\bfseries 1504}
  (2015) 0.25}, \href{http://arxiv.org/abs/1412.6302}{{\tt arXiv:1412.6302}} [\href{http://inspirehep.net/record/1335131}{Inspire}].

\bibitem{ATLAScollaboration:2014gl}
{\bfseries ATLAS} Collaboration, G.~Aad {\em et al.,}
  \href{http://dx.doi.org/10.1007/JHEP09(2014)037}{{\em JHEP} {\bfseries 09}
  (2014) 037}, \href{http://arxiv.org/abs/1407.7494}{{\tt arXiv:1407.7494}} [\href{http://inspirehep.net/record/1308524}{Inspire}].

\bibitem{CMSCollaboration:2014vq}
{\bfseries CMS} Collaboration, V.~Khachatryan
  \href{http://arxiv.org/abs/1408.2745}{{\tt arXiv:1408.2745}} [\href{http://inspirehep.net/record/1310653}{Inspire}].

\bibitem{Hisano:2015gna}
J.~Hisano, N.~Nagata, and Y.~Omura, \href{http://arxiv.org/abs/1506.03931}{{\tt
  arXiv:1506.03931}} [\href{http://inspirehep.net/record/1376004}{Inspire}].

\bibitem{Cheung:2015nha}
K.~Cheung, W.-Y. Keung, P.-Y. Tseng, and T.-C. Yuan,
  \href{http://arxiv.org/abs/1506.06064}{{\tt arXiv:1506.06064}} [\href{http://inspirehep.net/record/1377207}{Inspire}].

\bibitem{Dobrescu:2015qna}
B.~A. Dobrescu and Z.~Liu, \href{http://arxiv.org/abs/1506.06736}{{\tt
  arXiv:1506.06736}} [\href{http://inspirehep.net/record/1377366}{Inspire}].

\bibitem{Alves:2015mua}
A.~Alves, A.~Berlin, S.~Profumo, and F.~S. Queiroz,
  \href{http://arxiv.org/abs/1506.06767}{{\tt arXiv:1506.06767}} [\href{http://inspirehep.net/record/1377544}{Inspire}].

\bibitem{Gao:2015irw}
Y.~Gao, T.~Ghosh, K.~Sinha, and J.-H. Yu,
  \href{http://arxiv.org/abs/1506.07511}{{\tt arXiv:1506.07511}} [\href{http://inspirehep.net/record/1377754}{Inspire}].

\bibitem{Greco:2014aza}
D.~Greco and D.~Liu, \href{http://dx.doi.org/10.1007/JHEP12(2014)126}{{\em
  JHEP} {\bfseries 1412} (2014) 126},
  \href{http://arxiv.org/abs/1410.2883}{{\tt arXiv:1410.2883}} [\href{http://inspirehep.net/record/1321554}{Inspire}].

\end{thebibliography}\endgroup

\end{document}